\documentclass{cpbtex}
\usepackage{graphicx}
\usepackage{array}
\usepackage{subfig}
\usepackage[percent]{overpic}
\usepackage{bm}
\usepackage{mathdots}
\usepackage{booktabs}
\usepackage[font=small,labelfont=bf,labelsep=period]{caption}
\begin{document}
%\begin{CJK*}{GBK}{song}

\title{Majorana stellar representation for mixed-spin $(s, 1/2)$ systems\thanks{Project supported by the National Key Research and Development Program of China (No.~2017YFA0304202 and No.~2017YFA0205700), the NSFC through Grant No.~11875231, and the Fundamental Research Funds for the Central Universities through
Grant No.~2018FZA3005.}}

% Title should be concise; avoid abbreviations if possible; and not begin with `A', `An', `The', or `Study on'.

\author{Yu-Guo Su$^{1}$, \ Fei Yao$^{1}$, \ Hong-Bin Liang$^{1}$, \ Yan-Ming Che$^{1}$,\\ \ Li-Bin Fu$^{2}$, \ and \ Xiao-Guang Wang$^{1}$\thanks{Corresponding author. E-mail:xgwang1208@zju.edu.cn}\\
$^{1}${Zhejiang Province Key Laboratory of Quantum Technology and Device,}\\ {Department of Physics, Zhejiang University, Hangzhou, Zhejiang 310027, China}\\  % The line break was forced via \\
$^{2}${National Laboratory of Science and Technology on Computational Physics,}\\
{Institute of Applied Physics and Computational Mathematics, Beijing 100088, China}}   % The line break was forced via \\

% 1. For Chinese authors, the name in Chinese characters should also be given. For example, Gang Liu(Áõ¸Õ), Xiao-Ming Li(ÀîÏþÃ÷)
% 2. Please ensure that every author approves the submission of the manuscript
% 3. Abbreviations should not be used in the affiliations

\date{\today}
\maketitle

\begin{abstract}
By describing the evolution of a quantum state with the trajectories of the Majorana stars on a Bloch sphere, Majorana's stellar representation provides an intuitive geometric perspective to comprehend the quantum system with high-dimensional Hilbert space.
However, the problem of the representation of a two-spin coupling system on a Bloch sphere has not been solved satisfactorily yet.
Here, a practical method is presented to resolve the problem for the mixed-spin $(s, 1/2)$ system and describe the entanglement of the system.
The system can be decomposed into two spins: spin-$(s+1/2)$ and spin-$(s-1/2)$ at the coupling bases, which can be regarded as independent spins.
Besides, any pure state may be written as a superposition of two orthonormal states with one spin-$(s+1/2)$ state and the other spin-$(s-1/2)$ state.
Thus, the whole initial state can be regarded as a state of a pseudo spin-$1/2$.
In this way, the mixed spin decomposes into three spins. Therefore, the state can be represented by $(2s+1)+(2s-1)+1=4s+1$ sets of stars on a Bloch sphere.
Finally, some examples are given to show symmetric patterns on the Bloch sphere and unveil the properties of the high-spin system by analyzing the trajectories of the Majorana stars on the Bloch sphere.
\end{abstract}



\section{Introduction}\label{sec1}
It is acknowledged that the evolution of an arbitrary two-level state can be exactly represented by the trajectory of a point on the Bloch sphere.\ucite{Bloch1945,Niu2012,Schwinger1965} Applied to a quantum state in a high-dimensional Hilbert space, this geometric interpretation seems difficult to imagine.  Despite we can map the quantum pure state to a higher-dimensional geometric structure, this process is no more an intuitive and legible way to comprehend it. Fortunately, the Majorana's stellar representation (MSR) builds a wide bridge between the high dimensional projective Hilbert space and the two-dimensional Bloch sphere.\ucite{Majorana1932} Employing the MSR, which represents a quantum pure state of spin-$J$ systems in terms of a symmetrized state of 2$J$ spin-$1/2$ systems, one can generalize this geometric approach to large spin systems or multilevel systems. Majorana's perspective was that the evolution of a spin-$J$ state can be intuitively described by trajectories of 2$J$ points on the two-dimensional (2D) Bloch sphere, with these 2$J$ points generally coined as Majorana stars (MSs), rather than one point on an intricate high-dimensional geometric structure. Therefore, this representation spontaneously provides an intuitive way to study high spin systems from geometrical perspectives, which has made the MSR a useful tool in many different fields, e.g., classification of entanglement in symmetric quantum states,\ucite{Maekelae2010,Maekelae2010a,Ganczarek2012,Ribeiro2011,Markham2011,Mandilara2014,Guo2018,Liu2017CPB} analyzing the spectrum of the Lipkin-Meshkov-Glick model,\ucite{Ribeiro2007,Ribeiro2008} studying Bose condensate with high spins,\ucite{Barnett2006,Barnett2007,Yang2015,Stamper-Kurn2013,Kawaguchi2012,Lian2012,Cui2013,Liu2017APS} and calculating geometrical phases of large-spin systems.\ucite{Bruno2012,Liu2014,Niu2015} In addition, the MSR can be employed in quantum metrology\ucite{Bouchard2017,Goldberg2018} and in describing polarization states of $N$ photons.\ucite{Goldberg2017}

Moreover, the MSR provides many useful insights into high dimensional quantum states. The Berry phase, which is a unique character of a quantum state\ucite{Berry1984} and has become a central unifying concept for quantum state,\ucite{C2004,Bohm2003} unveils the gauge structure associated with cyclic evolution in Hilbert space.\ucite{Simon1983} When it comes to an arbitrary two-level state, the Berry phase is simply proportional to the solid angle subtended by the close trajectory of a point on the Bloch sphere, while every star in the MSR will trace out its own trajectory on the Bloch sphere for a cyclic evolution of a large spin state. For example, the Majorana stars can be driven moving periodically on the Bloch sphere, and making up the so-called ``Majorana spin helix''\ucite{Cui2013} by the spin-orbit coupling in high-spin condensates. Consequently, by asking what the explicit relation between the Berry phase and the Majorana stars' helixes or loops is, it has turned into a significant topic in recent years.\ucite{Hannay1998,Liu2014,Bruno2012,Tamate2011,Ogawa2015}
% Hence, one of the insights the MSR brought for a quantum state is how to visualize the Berry phase of a large spin state by the trajectories of stars on the Bloch sphere such as the solid angle for the spin-$1/2$ state.

Except for the Berry phase, entanglement is another important unique character of a many-particle quantum state. Although the classification and measure are quite complex\ucite{PhysRevA.61.052306,PhysRevA.67.012108,PhysRevA.62.062314,PhysRevLett.87.040401,Mandilara2014} for the multiqubit states, the MSR naturally provides an intuitive way to consider the multiqubit entanglement,\ucite{Liu2016} since a spin-$J$ state is equivalent to a symmetric 2$J$-qubit state. The distribution of the Majorana stars not only discloses the relationship between the symmetry of the state and the multiparticle entanglement measures, including geometric measures\ucite{Aulbach2010,Markham2011,PhysRevLett.108.210407,PhysRevA.87.012104} and Barycentric measures,\ucite{Ganczarek2012} but also can be employed to investigate entanglement classes,\ucite{PhysRevLett.103.070503,PhysRevA.81.052315} entanglement invariants,\ucite{PhysRevLett.106.180502} and so on. Hence, it is another challenge task to connect the quantum entanglement of the qubits to the distribution of the Majorana stars on the Bloch sphere.

Furthermore, it is interesting to apply this approach to study the multiband topological systems.
%It is well known that a topological insulator distinguishes a trivial band insulator by its nontrivial topological energy band, which has different geometrical property from a trivial band [18].
For a two-band system, e.g., the Su-Schrieffer-Heeger (SSH) model,\ucite{PhysRevLett.42.1698,PhysRevB.22.2099,Su1981} the geometrical meaning of topologically different phases can be revealed by their distinct trajectories,\ucite{PhysRevLett.89.077002,PhysRevB.84.195452} with mapping the Bloch state into a 2D Bloch sphere. As a paragon topological model,\ucite{Shen} the SSH model supports either topologically trivial or nontrivial phase, characterized by the quantized Berry phase 0 or $\pi$,\ucite{Berry1984,PhysRevLett.62.2747,RevModPhys.82.1959} which is verified in the recent cold atom experiment.\ucite{Atala2013}

From Refs.~\cite{Majorana1932,Liu2017,Yao2017}, an arbitrary pure state for spin $J$ can be represented by
\begin{eqnarray}
  |\psi\rangle_{j} & = &  \sum^{2j}_{n=0}C_j^{(n)}|n\rangle_j, \label{pure_psi}
\end{eqnarray}
where $j$ is the angular quantum number, $|n\rangle_j\equiv |-j+n\rangle_j$ is the basis state and $C_j^{(n)}$ is the corresponding coefficient. The star equation is given by\ucite{Majorana1932}
\begin{eqnarray}
\sum_{n=0}^{2j}(-1)^{n}\left(\begin{array}{c}
2j\\
n
\end{array}\right)^{\frac{1}{2}}C_j^{(n)}z^{n}=0,\label{C_n&z}
\end{eqnarray}
where $z$ denotes the characteristic variable and we also have the normalization condition $\sum_{n=0}^{2j}|C^{(j)}_{n}|^2=1$. The root $z$ can be mapped to the stars on Bloch sphere via relation
\begin{eqnarray}
  z=\tan\frac{\theta}{2}\text{e}^{\text{i}\phi},\theta\in[0,\pi],\phi\in[0,2\pi],\label{z_tan&e}
\end{eqnarray}
where $\theta$ and $\phi$ are the spherical coordinates.

The work of Brody and Hughston\ucite{Brody1998,Brody2001} gives an intuitive formalism around higher-dimensional geometries of pure states. They formulated the principles of classical statistical inference in a natural geometric setting and gave a number of examples of features in the state spaces for higher-dimensional systems.
The two-spin coupling system (a large spin coupling with a small spin) is significant in many physical fields, such as quantum criticality\ucite{Quan2006} and quantum dynamics.\ucite{Ahmadi2010}
However, the representation of a two-spin coupling system has been studied fragmentarily,\ucite{Fontana1962,DeRaedt1983,Haaker2014,Bjoerk2015} which did not generalize to an arbitrary spin-$s$ and was not visible enough.
In this work, we choose mixed-spin $(s, 1/2)$ systems as paragons to take advantage of the fact that it can be decomposed into two spins: spin-$(s+1/2)$ and spin-$(s-1/2)$. Benefiting from the MSR, we represent an arbitrary pure state on a Bloch sphere. Besides, we propose a practical method to decompose the arbitrary pure state that can be regarded as a state of a pseudo spin-$1/2$.
%For the arbitrary pure state, we always can rewrite it as a superposition of two pure-orthonormal states:
%\begin{align}
%  |\psi\rangle_{j=s+\frac{1}{2}}=C_{s,\frac{1}{2}}^+|\psi\rangle_{s+\frac{1}{2}}+C_{s,\frac{1}{2}}^-|\psi\rangle_{s-\frac{1}{2}},\label{PseudoSpin1/2_1/2}
%\end{align}
%where $|\psi\rangle_{s+\frac{1}{2}}$ and$|\psi\rangle_{s-\frac{1}{2}}$ are spin-($s+1/2$) normalized state and spin-($s-1/2$) normalized state, respectively, and $C_{s,1/2}^+$ and $C_{s,1/2}^-$ are the normalized coefficients with $|C_{s,1/2}^+|^2+|C_{s,1/2}^-|^2=1$.
For the arbitrary pure state of mixed spin, the system can be decomposed into two spins: spin-$(s+1/2)$ and spin-$(s-1/2)$.
Consequently, the arbitrary pure state of the mixed spin can be regarded as a state of a pseudo spin-$1/2$.
%From \cref{pure_psi} and Refs.\ucite{Liu2017,Yao2017}, we obtain the general expression of the analogous triplet and the analogous singlet
%\begin{align}
%  |\psi\rangle^{j} &=\sum_{n=0}^{2j}(-1)^{n}\binom{2j}{n}^{1/2}C_n^{(j)}(-\tilde{J}_{+})^{n}|0\rangle_j,
%\end{align}
%where the angular quantum number $j=2s\pm1$ are the analogous triplet case (plus) and the analogous singlet case (minus), $|0\rangle_j$ are their ground state, $C_{n}^{(j)}$ denote the corresponding coefficients of the eigenstates and we introduce the nonlinear creation operator $\tilde{J}_{+}$.
In this way, the mixed spin decomposes into three spins and our task is resolving these star equations and obtaining $(2s+1)+(2s-1)+1=4s+1$ sets of stars.

Our study provides an intuitive perspective of a two-spin ($s$ and $1/2$) system and unveils the intrinsic property of the two-spin system on a Bloch sphere, which shall deepen our comprehension of the spin-$(s, 1/2)$ system. The paper is organized as follows. In Section~2, we study the fundamental theory of the MSR to describe an arbitrary pure state of spin-$(s, 1/2)$, through coupling bases. In Section~3, we give a concise example for the MSR of the mixed-spin $(1/2, 1/2)$ systems. In Section~4, we show more applications of our method in the mixed-spin $(s, 1/2)$ systems. A brief discussion and summary are given in Section~5.

\section{Theory of Majorana representation for mixed-spin (s,1/2) systems}\label{sec2}
In this section, we will study the fundamental theory of the MSR for the mixed-spin $(s, 1/2)$ systems. Firstly, we illustrate with the mixed spin-$(1/2, 1/2)$ system to give a legible view of our main logic. Besides, we introduce the coupling bases to present an arbitrary mixed spin-$(s, 1/2)$ state as two independent spins.
Finally, benefiting from the new form, a two-level system can be constructed to describe the arbitrary pure state.

For two spin-$1/2$ particles, an arbitrary pure state can be represented by
\begin{eqnarray}
  |\psi\rangle_{\frac{1}{2},\frac{1}{2}}=a\left|\uparrow\uparrow\right\rangle+b\left|\downarrow\uparrow\right\rangle+c\left|\uparrow\downarrow\right\rangle+d\left|\downarrow\downarrow\right\rangle. \label{1/2&1/2_psi}
\end{eqnarray}
We can rewrite Eq.~(\ref{1/2&1/2_psi})
\begin{eqnarray}
  |\psi\rangle_{\frac{1}{2},\frac{1}{2}}
  =\sqrt{\frac{2-|b-c|^2}{2}}\left|\Uparrow\right\rangle+\frac{-b+c}{\sqrt{2}}\left|\Downarrow\right\rangle,\label{triplet&singlet}
\end{eqnarray}
where %$\left|\Uparrow\right\rangle=[a\left|\uparrow\uparrow\right\rangle+d\left|\downarrow\downarrow\right\rangle+(b+c)/2(\left|\uparrow\downarrow\right\rangle+\left|\downarrow\uparrow\right\rangle)]/\sqrt{|a|^2+|d|^2+|b+c|^2/2}$,
%$\left|\Uparrow\right\rangle=[a\left|\uparrow\uparrow\right\rangle+d\left|\downarrow\downarrow\right\rangle+(b+c)(\left|\uparrow\downarrow\right\rangle+\left|\downarrow\uparrow\right\rangle)/2]/\sqrt{1-|b-c|^2/2}$, $\left|\Downarrow\right\rangle = (\left|\uparrow\downarrow\right\rangle-\left|\downarrow\uparrow\right\rangle)/\sqrt{2}$.
%$\left|\Uparrow\right\rangle=\frac{a\left|\uparrow\uparrow\right\rangle+d\left|\downarrow\downarrow\right\rangle+\frac{(b+c)(\left|\uparrow\downarrow\right\rangle+\left|\downarrow\uparrow\right\rangle)}{2}}{\sqrt{\frac{2-|b-c|^2}{2}}}$,
$\left|\Uparrow\right\rangle=\left[a\left|\uparrow\uparrow\right\rangle+d\left|\downarrow\downarrow\right\rangle+(b+c)(\left|\uparrow\downarrow\right\rangle+\left|\downarrow\uparrow\right\rangle)/2\right]/\sqrt{\left(2-|b-c|^2\right)/2}$,  $\left|\Downarrow\right\rangle = \left(\left|\uparrow\downarrow\right\rangle-\left|\downarrow\uparrow\right\rangle\right)/\sqrt{2}$.
Then we notice that the terms of the bracket are the analogous triplet $|\psi\rangle^{j=1}$ and the last term is the analogous singlet $|\psi\rangle^{j=0}$.

For more general case, we can treat the system as a two-level energy system, which can be represented as two sets of stars on a Bloch sphere if it is a pure state.
To normalize the states, we rewrite Eq.~(\ref{triplet&singlet}) as
\begin{eqnarray}
|\psi\rangle=C_{s+\frac{1}{2}}|\psi\rangle_{s+\frac{1}{2}}+C_{s-\frac{1}{2}}|\psi\rangle_{s-\frac{1}{2}},\label{correlation_point1}
\end{eqnarray}
where $|\psi\rangle_{s+1/2}$ and $|\psi\rangle_{s-1/2}$ represent the normalized analogous triplet and the normalized analogous singlet.
We write the whole initial state as a superposition of two orthogonal states $|\psi\rangle_{s+1/2}$ and $|\psi\rangle_{s-1/2}$.
We also have the normalization condition $|C_{s+1/2}|^2+|C_{s-1/2}|^2=1$, so we only have one independent variable to describe the relation
%also have the normalization condition $|C_{s+1/2}|^2+|C_{s-1/2}|^2=1$, so we only have one independent variable to describe the relation
between them, which means that it needs one set of stars on the Bloch sphere to be represented.
Based on Eqs.~(\ref{C_n&z}) and~(\ref{correlation_point1}), we can treat the state as a pseudo spin-$1/2$ and have the star (root) of the pseudo spin
\begin{eqnarray}
  z_{s,\frac{1}{2}}=\frac{C_{s-\frac{1}{2}}}{C_{s+\frac{1}{2}}},\label{z_j=1/2}
\end{eqnarray}
which reveals the relation between the analogous triplet and the analogous singlet.
Because the orthogonality of $|\psi\rangle_{s+1/2}$ and $|\psi\rangle_{s-1/2}$ do not constrain the phase difference between them, we choose zero phase difference and keep the star of the pseudo spin on the prime meridian in MSR.
Noticing the completeness condition, if one obviates the state only has components in some parts of the whole sectors, one find that the mapping is injective.
Therefore, we need $(2s+1)$ sets of stars to represent the analogous triplet, $(2s-1)$ sets of stars to represent the analogous singlet and one set of  stars to combine the two parts on a Bloch sphere.

From Eqs.~(\ref{pure_psi}) and~(\ref{C_n&z}), we can get all stars %(calculation details see Appendix B)
\begin{eqnarray}
z_1^{(\pm)}&=\frac{(b+c)\pm \sqrt{(b+c)^2-4ad}}{2a},\label{1/2&1/2_z_1}\\
z_{\frac{1}{2}}^{(1)}&=\frac{b-c}{\sqrt{2-\left|b-c\right|^2}},\label{1/2&1/2_z}
\end{eqnarray}
where $z_1^{(\pm)}$ and $z_{1/2}^{(1)}$ are respective roots of the analogous triplet (singlet) and the state of the pseudo spin. Therefore, we
%where $z_1^{(\pm)}$ and $z_{1/2}^{(1)}$ are respective roots of the analogous triplet and the analogous state of the pseudo spin. Therefore, we
need two sets of stars to represent the analogous triplet, no stars to represent the analogous singlet and one set of  stars to combine the two parts on a Bloch sphere.

Now, we show a general method to gain the construct of the pseudo spin-1/2. For the mixed-spin $(s, 1/2)$ systems, we have the total angular momentum ${\bm J}={\bm S}+{\bm \sigma}/2$ (we choose $\hbar$=1), and ${\bm S}$, ${\bm \sigma}/2$ are the angular momentums of the large spin-$s$ and small spin-$1/2$, respectively.
Therefore, we have
\begin{eqnarray}
{\bm J}^2&=&{\bm S}^2+\frac{1}{4}{\bm \sigma}^2+{\bm S}_{z}\sigma_{z}+({\bm S}_{+}\sigma_{-}+{\bm S}_{-}\sigma_{+}),\label{J}\\
{\bm S}_{+}|n\rangle_{s}&=&\sqrt{(n+1)(2s-n)}|n+1\rangle_{s},\label{S+}\\
{\bm S}_{-}|n\rangle_{s}&=&\sqrt{n(2s-n+1)}|n-1\rangle_{s},\label{S-}\\
{\bm S}_{z}|n\rangle_{s}&=&(-s+n)|n\rangle_{s},\label{Sz}
\end{eqnarray}
where ${\bm S}={\bm S}_{x}+{\bm S}_{y}+{\bm S}_{z}$ is the angular momentum, ${\bm S}_{+}={\bm S}_{x}\pm \text{i}{\bm S}_{y}$ is the raising (lowering) operator  and $|n\rangle_{s}\equiv |-s+n\rangle_{s}$ is the basis state.
Besides, we write the uncoupling basis state as $\left|n\right\rangle_{s}{\bm \otimes}|1\rangle_{1/2}\equiv|n\rangle_1|1\rangle_2$, $\left|n\right\rangle_{s}{\bm \otimes}|0\rangle_{1/2}\equiv|n\rangle_1|0\rangle_2$, and $n=0,1,\ldots,2s$.

Then we have the Hamiltonian
\begin{eqnarray}
  H={\bm J}^2, \label{H:J}
\end{eqnarray}
and its vector space $W$.
We find that the space $V_n$ is the two-dimensional invariant subspace of the vector space $W={\bm \oplus}^{2s}_{n=1}V_n{\bm \oplus} V_0{\bm \oplus} V_{2s+1}$, where $V_0$ and $V_{2s+1}$ are one-dimensional subspaces.
The bases of the subspace $V_n$ $\left(1\leq n\leq 2s\right)$ are $|n\rangle_1|0\rangle_2$ and $\left|n-1\right\rangle_1|1\rangle_2$.
The bases of the subspace $V_0$ and $V_{2s+1}$ are $|0\rangle_1|0\rangle_2$ and $|2s\rangle_1|1\rangle_2$, respectively.

From Eqs.~(\ref{J})-(\ref{H:J}), we can rewrite the Hamiltonian as
%From \cref{J,S+,S-,Sz,H:J}, We can rewrite the Hamiltonian as
\begin{eqnarray}
  H={\bm \oplus}^{2s+1}_{n=0}A_n,
\end{eqnarray}
where the two-dimensional matrix
\begin{eqnarray}
  A_n=(s^2+s+\frac{1}{4})I+(s-n+\frac{1}{2})\sigma_z+\sqrt{n(2s-n+1)}\sigma_x \label{A_n}
\end{eqnarray}
for $n=1,\ldots,2s$,
$A_{0}=A_{2s+1}=(s^2+2s+3/4)$.
From Eq.~(\ref{A_n}), we diagonalize $A_n$ and then get the eigenvalue $j_{\pm}(j_{\pm}+1)$ ($j_{\pm}=s\pm1/2$) of the $A_n$, and the relationship between the coupling and the uncoupling representations
%(see Appendix A for detailed calculations)
\begin{eqnarray}
  |n\rangle_{s+\frac{1}{2}} &=\sqrt{\frac{2s-n+1}{2s+1}}|n\rangle_1|0\rangle_2+\sqrt{\frac{n}{2s+1}}|n-1\rangle_1|1\rangle_2,\label{s+1/2}\\
  |n\rangle_{s-\frac{1}{2}} &=\sqrt{\frac{n}{2s+1}}|n\rangle_1|0\rangle_2-\sqrt{\frac{2s-n+1}{2s+1}}|n-1\rangle_1|1\rangle_2,\label{s-1/2}
\end{eqnarray}
where $n=1,\ldots,2s$, $|n\rangle_{s+1/2}$ and $|n\rangle_{s-1/2}$ denote the coupling basis states.

From Eqs.~(\ref{s+1/2}) and~(\ref{s-1/2}), we can get the uncoupling basis states
%From \cref{s+1/2,s-1/2}, we can get the uncoupling basis states
\begin{eqnarray}
  |n\rangle_1|0\rangle_2 &=\sqrt{\frac{2s-n+1}{2s+1}}|n\rangle_{s+\frac{1}{2}}+\sqrt{\frac{n}{2s+1}}|n\rangle_{s-\frac{1}{2}},\label{n_0}\\
  |n-1\rangle_1|1\rangle_2 &=\sqrt{\frac{n}{2s+1}}|n\rangle_{s+\frac{1}{2}}-\sqrt{\frac{2s-n+1}{2s+1}}|n\rangle_{s-\frac{1}{2}},\label{n-1_1}
\end{eqnarray}
where $n=1,\ldots,2s$.

For a mixed spin-$(s, 1/2)$ system, using Eqs.~(\ref{n_0}) and~(\ref{n-1_1}), we can rewrite the arbitrary state
%For a mixed spin-$(s, 1/2)$ system, using \cref{n_0,n-1_1}, we can rewrite the arbitrary state
%(see Appendix A for detailed calculations)
\begin{eqnarray}
% |\psi\rangle&=&\sum^{2s}_{m=0}{D_{m,1}|m\rangle_1|1\rangle_2}+\sum^{2s}_{n=0}{D_{n,0}|n\rangle_1|0\rangle_2}\nonumber\\
%   &=&D_{0,0}|0\rangle_{1}|0\rangle_{2}+D_{2s,1}|2s\rangle_1|1\rangle_2\nonumber\\
%   &&+\sum^{2s}_{n=1}{\left(D_{n-1,1}|n-1\rangle_1|1\rangle_2+D_{n,0}|n\rangle_1|0\rangle_2\right)}\nonumber\\
%  &=&D_{0,0}|0\rangle_{1}|0\rangle_{2}+D_{2s,1}|2s\rangle_1|1\rangle_2\nonumber\\
%  &&+\sum^{2s}_{n=1}{\left(E_{n,s+\frac{1}{2}}|n\rangle_{s+\frac{1}{2}}+F_{n,s-\frac{1}{2}}|n\rangle_{s-\frac{1}{2}}\right)}\nonumber\\
%  &=&\sum^{2s+1}_{n=0}{E_{n,s+\frac{1}{2}}|n\rangle_{s+\frac{1}{2}}}+\sum^{2s}_{n=1}{F_{n,s-\frac{1}{2}}|n\rangle_{s-\frac{1}{2}}},\label{s&1/2_psi}
 |\psi\rangle&=&\sum^{2s}_{m=0}{D_{m,1}|m\rangle_1|1\rangle_2}+\sum^{2s}_{n=0}{D_{n,0}|n\rangle_1|0\rangle_2}\nonumber\\
   &=&D_{0,0}|0\rangle_{1}|0\rangle_{2}+D_{2s,1}|2s\rangle_1|1\rangle_2+\sum^{2s}_{n=1}{\left(D_{n-1,1}|n-1\rangle_1|1\rangle_2+D_{n,0}|n\rangle_1|0\rangle_2\right)}\nonumber\\
  &=&D_{0,0}|0\rangle_{1}|0\rangle_{2}+D_{2s,1}|2s\rangle_1|1\rangle_2+\sum^{2s}_{n=1}{\left(E_{n,s+\frac{1}{2}}|n\rangle_{s+\frac{1}{2}}+F_{n,s-\frac{1}{2}}|n\rangle_{s-\frac{1}{2}}\right)}\nonumber\\
  &=&\sum^{2s+1}_{n=0}{E_{n,s+\frac{1}{2}}|n\rangle_{s+\frac{1}{2}}}+\sum^{2s}_{n=1}{F_{n,s-\frac{1}{2}}|n\rangle_{s-\frac{1}{2}}},\label{s&1/2_psi}
\end{eqnarray}
where
\begin{eqnarray}
  E_{n,s+\frac{1}{2}}&=\frac{D_{n-1,1}\sqrt{n}+D_{n,0}\sqrt{2s-n+1}}{\sqrt{2s+1}},\\
  F_{n,s-\frac{1}{2}}&=\frac{-D_{n-1,1}\sqrt{2s-n+1}+D_{n,0}\sqrt{n}}{\sqrt{2s+1}},
\end{eqnarray}
and $n=m+1$, $D_{m,1}=D_{n-1,1}$, $D_{n,0}$ are the coefficients of the eigenstates, respectively. We notice that the terms of the bracket are the analogous multiplet and the last term is the analogous singlet.

\section{Application in mixed-spin $(1/2, 1/2)$ systems with a real phase parameter $\varphi$ and time evolution}\label{sec3}

In this section, we will show a concise example for the mixed-spin $(s, 1/2)$ systems with a real phase parameter $\varphi$.

From Eqs.~(\ref{z_tan&e}),~(\ref{1/2&1/2_z_1}) and~(\ref{1/2&1/2_z}), we use two sets of stars to represent the analogous triplet, no star to represent the analogous
%From \cref{1/2&1/2_z_1,1/2&1/2_z,z_tan&e}, we use two sets of stars to represent the analogous triplet, no star to represent the analogous
singlet and one set of stars to combine the two parts on a Bloch sphere.

To illustrate the idea above, we give a simple spin-$(1/2, 1/2)$ system with time evolution:
\begin{eqnarray}
%|\psi\rangle_{\frac{1}{2},\frac{1}{2}}&=&e^{-{\rm i} H_{1}t}\left(\cos\varphi\left|\uparrow\right\rangle+\sin\varphi\left|\downarrow\right\rangle\right)\nonumber\\
%&&{\bm \otimes}\left(\sin\varphi\left|\uparrow\right\rangle+\cos\varphi\left|\downarrow\right\rangle\right),
|\psi\rangle_{\frac{1}{2},\frac{1}{2}}&=&\text{e}^{-{\rm i} H_{1}t}\left(\cos\varphi\left|\uparrow\right\rangle+\sin\varphi\left|\downarrow\right\rangle\right){\bm \otimes}\left(\sin\varphi\left|\uparrow\right\rangle+\cos\varphi\left|\downarrow\right\rangle\right),
\end{eqnarray}
where the Hamiltonian $H_{1}=\sigma_{1x}\sigma_{2x}+\sigma_{1y}\sigma_{2y}+\delta\sigma_{1z}\sigma_{2z}$, $\delta\in[0,1]$ and the real phase parameter $\varphi\in(0,\pi/2)\cup(\pi/2,\pi)\cup(\pi,3\pi/2)\cup(3\pi/2,2\pi)$, considering the completeness condition.

\begin{figure}[htbp]
\centering
\begin{minipage}{1\linewidth}
\centering
\begin{overpic}[width=0.667\linewidth,height=0.667\linewidth]{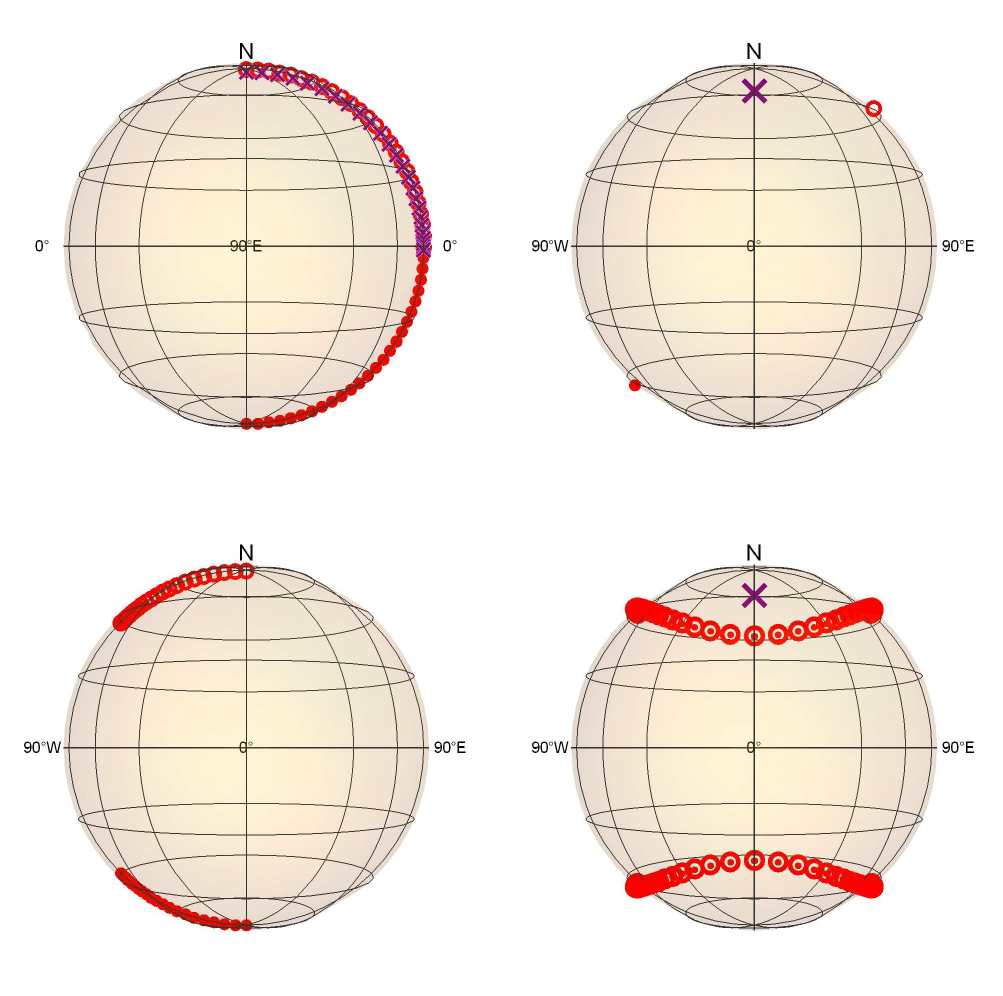}%\label{fig3-a}
\put(0,90){(a)}
\put(50,90){(b)}
\put(0,45){(c)}
\put(50,45){(d)}
\end{overpic}
\end{minipage}
\protect\caption{\label{fig1}(color online) Bloch representation of the two spin-$1/2$ particles with the real phase parameter $\varphi$, time evolution and $\delta=0$. $a$-$c$) The stars are mapped to the Bloch sphere with the variety of $\varphi$, meanwhile, without time evolution. a) All stars with $t=0$. b) All stars with $t=\pi/6$ and $\varphi=\pi/3$. c) The triplet stars with $t=\pi/4$. d) All stars are mapped to the Bloch sphere with the period of time evolution $T=\pi$ and $\varphi=2\pi/3$. The red circles and red dots represent the stars of the triplet, purple crosses represent the stars of the pseudo spin.
}
\end{figure}

Therefore, we gain triplet stars
\begin{eqnarray}
0&=&z_{1}^2 e^{-{\rm i} \delta  t} \sin (2\varphi )-2z_{1}e^{{\rm i} (\delta -2) t}+e^{-{\rm i} \delta  t} \sin (2\varphi ),\\
%z_{1}=\tan(\frac{\theta_{1,2}}{2})e^{i\phi_{1,2}},\\
\pi&=&\theta_++\theta_-,\\
0&=&\phi_++\phi_-,\\
%\phi^{\prime}_+-\phi^{\prime}_-=\pi,
\theta_{\pm}&=&2 \arctan\left|\left(e^{-2{\rm i} t(\delta-1)}\pm\frac{\sqrt{2}}{2}\lambda\right)\csc(2\varphi)\right|,\\
\phi_{\pm}&=&\arctan\left(\frac{{\rm Im}\left[\left(e^{-2{\rm i} t(\delta-1)}\pm\frac{\sqrt{2}}{2} \lambda\right)\csc (2\varphi )\right]}{{\rm Re}\left[\left(e^{-2{\rm i} t(\delta-1)}\pm\frac{\sqrt{2}}{2}\lambda\right) \csc (2\varphi )\right]}\right),
%\phi^{\prime}_{\pm}=\sin ^{-1}\big(\frac{\rm{Im}((e^{-2 it(\delta-1)}\pm\frac{\sqrt{2}}{2} \sqrt{2 e^{-4 it(\delta-1)}+\cos (4 \varphi )-1})\csc (2\varphi ))}{| (e^{-2 it(\delta-1)}\pm\frac{\sqrt{2}}{2}\sqrt{\cos (4 \varphi )+2 e^{-4 it(\delta-1)}-1}) \csc (2\varphi )| }\big).
\end{eqnarray}
where $\lambda=\sqrt{\cos(4\varphi)+2e^{-4{\rm i} t(\delta-1)}-1}$, $\theta_{\pm}$ and $\phi_{\pm}$ are the zenith angle and azimuth angle of the triplet stars. Obviously, we can find that the two sets of the stars are $180^{\circ}$ rotational symmetric around the $x$-axis (the intersecting line of the prime meridian plane and the equatorial plane).

However, singlet state has no star. The stars (roots) of the pseudo spin
\begin{eqnarray}
z_{\frac{1}{2},\frac{1}{2}}&=&\sqrt{\frac{\cos^2(2 \varphi )}{\sin ^2(2 \varphi )+1}},\\
\theta&=&2 \arctan\left(\sqrt{\frac{\cos^2(2 \varphi )}{\sin ^2(2 \varphi )+1}}\right)\in[0,\frac{\pi}{2}],\\
\phi&=&0.%,\\
%\phi^{\prime}&=0.
\end{eqnarray}

Hence, we know that the stars of the pseudo spin are always mapped to the north of the prime meridian and they are independent on time evolution. To show the idea above, we give a brief example of the two spin-$1/2$ particles with the real phase parameter $\varphi$, time evolution and $\delta=0$.

Figure \ref{fig1} shows the stars of the two spin-$1/2$ particles represented in Bloch sphere with real phase parameter $\varphi$, time evolution and $\delta=0$. Without time evolution, because the roots ($z_1$) in our example are always real, with the variety of the real phase parameter $\varphi$, each set of the triplet stars is mapped to the north or south prime meridian. Intuitively, the stars of the pseudo spin only cover the north prime meridian.
Since star of the analogous singlet does not have root in two spin-$1/2$ case, we don't have singlet state star (Fig.~\ref{fig1}(a)).
Obviously, the position of the star of pseudo spin unveils the relationship of corresponding coefficients between $|\psi\rangle_{s+1/2}$ and $|\psi\rangle_{s-1/2}$.
For example, if the latitude of the star of pseudo spin rises and the stars of pseudo spin are sparse, it means that the proportion of $|\psi\rangle_{s-1/2}$ increases and speed of the change is high.
Fixed the time at a special value, we can distinctly find that one set of the triplet stars and the others are $180^{\circ}$ rotational symmetric around the $x$-axis (the intersecting line of the prime meridian plane and the equatorial plane) (Fig.~\ref{fig1}(b)).
As seen in Fig.~\ref{fig1}(c), the triplet stars are mapped to the prime meridian at $t=2k\pi/4$ , the $90^{\circ}W(E)$ meridian at $t=(2k+1)\pi/4$ ($k$ is integer).
%And the triplet stars are mapped to the prime meridian at $t=2k\pi/4$ , the $90^{\circ}W(E)$ meridian at $t=(2k+1)\pi/4$ ($k$ is integer).
Fixed the phase parameter $\varphi$ at a special value, since stars of the pseudo spin are independent of time evolution, they are fixed at particular positions (Fig.~\ref{fig1}(d)). Meanwhile, each set of the triplet stars is plane symmetry around the equatorial plane and the prime meridian plane at a whole period $T=\pi$, and each set still keeps the rotational symmetry at every single time and phase parameter value.

\begin{table*}[htbp]
	\centering
	\caption{Specific mixed-spin ($\frac{1}{2}$, $\frac{1}{2}$) systems in Majorana's stellar representation.}
	\begin{tabular}{cp{4cm}<{\centering}cp{3cm}<{\centering}cp{3cm}<{\centering}cp{3cm}<{\centering}}
		\toprule[1pt]
\specialrule{0em}{2pt}{2pt}
		\multicolumn{2}{c}{State}&Sector&Star&Property\\
\specialrule{0em}{2pt}{2pt}	
\midrule[1pt]
\specialrule{0em}{2pt}{2pt}
&$\left|\uparrow\uparrow\right\rangle$&+$^{\star}$&2&$S_+^{\circ}$\\
\specialrule{0em}{2pt}{2pt}
Non-entangled&$\left|\downarrow\downarrow\right\rangle$&+&0&None\\
\specialrule{0em}{2pt}{2pt}
&$\left|\downarrow\uparrow\right\rangle/\left|\uparrow\downarrow\right\rangle$&+ and -&2&$S_+$ and $E^\bullet$\\\cmidrule[1pt]{1-1}%\cmidrule(r){1-1}
\specialrule{0em}{2pt}{2pt}
&$\frac{1}{\sqrt{2}}\left(\left|\downarrow\uparrow\right\rangle\pm\left|\uparrow\downarrow\right\rangle\right)$&+ (-)&1 (0)&$S_+$ (None)\\
\specialrule{0em}{2pt}{2pt}
Entangled&$\frac{1}{\sqrt{2}}\left(\left|\uparrow\uparrow\right\rangle\pm\left|\downarrow\downarrow\right\rangle\right)$&+&2&Equatorial symmetry\\
\specialrule{0em}{2pt}{2pt}
&Others&---&---&Complex$^\Diamond$\\
\specialrule{0em}{2pt}{2pt}
		\bottomrule[1pt]
	\end{tabular}\label{Table}\\
\footnotesize{$^\star$ + (-) means the star is in the spin-($s+1/2$) (spin-($s-1/2$)) sector.}\\
\footnotesize{$^\circ$ $S_+$ means the star is on the south pole in the spin-($s+1/2$) sector.}\\
\footnotesize{$^\bullet$ $E$ means the star of the pseudo spin is on the equator.}\\
\footnotesize{$^\Diamond$ ``Complex'' means the stars' distribution is complex.}
\end{table*}

The method we proposed can visualize the mixed-spin ($s$, $1/2$) system and describe the entanglement of the system.
For the mixed-spin ($1/2$, $1/2$) system, the distributions of specific states' stars in Majorana's stellar representation are given in Table~\ref{Table}.
%In the mixed-spin ($\frac{1}{2}$, $\frac{1}{2}$) system, for example, $\left|\uparrow\uparrow\right\rangle$ ($\left|\downarrow\downarrow\right\rangle$) only belongs to the spin-$s+\frac{1}{2}$ sector, which means it is a non-entangled state.
%While, $\left|\downarrow\uparrow\right\rangle$ ($\left|\uparrow\downarrow\right\rangle$) exists on both sectors
%belongs to both sectors because it cannot be separated into one sector.
From Table~\ref{Table}, $\left|\uparrow\uparrow\right\rangle$ ($\left|\downarrow\downarrow\right\rangle$) only belongs to the spin-($s+1/2$) sector.
Although the stars of $\left|\downarrow\uparrow\right\rangle$ ($\left|\uparrow\downarrow\right\rangle$) belong to both sectors and it cannot be separated into one sector, they are respective on the south pole and the equator in MSR.
The Bell states
%$\left|\Psi^\pm\right\rangle=\frac{1}{\sqrt{2}}\left(\left|01\right\rangle\pm\left|01\right\rangle\right)$ and $\left|\Phi^\pm\right\rangle=\frac{1}{\sqrt{2}}\left(\left|00\right\rangle\pm\left|11\right\rangle\right)$
$\left|\Psi^\pm\right\rangle=\left(\left|\uparrow\downarrow\right\rangle\pm\left|\downarrow\uparrow\right\rangle\right)/\sqrt{2}$ and
$\left|\Phi^\pm\right\rangle=\left(\left|\uparrow\uparrow\right\rangle\pm\left|\downarrow\downarrow\right\rangle\right)/\sqrt{2}$
 only belong to one sector.
However, the stars' distributions of other entangled states are very complex.
Therefore, it is easy to find that all the non-entangled states' stars are on the south pole and the equator in MSR, so are the Bell states. But the distributions of general entangled states' are complex.

\section{Application in mixed-spin $(s, 1/2)$ systems with a real phase parameter $\varphi$ and time evolution}\label{sec4}
In this section, we will give brief examples with a real phase parameter $\varphi$, time fixed case, and with time evolution, fixed real phase parameter $\varphi$ case for the mixed- spin $(s, 1/2)$ systems, respectively. To illustrate the idea above, let's consider the example spin-$1$ and spin-$1/2$ case as
\begin{eqnarray}
%  |\psi\rangle_{1,\frac{1}{2}}&=&e^{-{\rm i} H_{2}t}\left(\frac{\cos\varphi}{\sqrt{2}}|1\rangle+\frac{1}{\sqrt{2}}|0\rangle+\frac{\sin\varphi}{\sqrt{2}}|-1\rangle\right)\nonumber\\
%  &&{\bm \otimes}\left(\cos\varphi\left|\uparrow\right\rangle+\sin\varphi\left|\downarrow\right\rangle\right),\label{1&1/2_psi}
  |\psi\rangle_{1,\frac{1}{2}}&=&\text{e}^{-{\rm i} H_{2}t}\left(\frac{\cos\varphi}{\sqrt{2}}|1\rangle+\frac{1}{\sqrt{2}}|0\rangle+\frac{\sin\varphi}{\sqrt{2}}|-1\rangle\right){\bm \otimes}\left(\cos\varphi\left|\uparrow\right\rangle+\sin\varphi\left|\downarrow\right\rangle\right),\label{1&1/2_psi}
\end{eqnarray}
where the Hamiltonian $H_{2}=S_{1x}S_{2x}+S_{1y}S_{2y}+\delta S_{1z}S_{2z}$, $S_{1i}$ (respectively, $S_{2i}$, $i=x,y,z$) are the three direction components of  the larger spin (respectively, spin-$1/2$), $|1\rangle$, $|0\rangle$, $|-1\rangle$ are the eigenstate of the spin-1 particle, and $\left|\uparrow\right\rangle$, $\left|\downarrow\right\rangle$ are the eigenstate of the spin-$1/2$ particle. So we have unitary transformation operator $U=\exp{(-{\rm i} H_{2}t)}$.

Utilizing Eqs.~(\ref{pure_psi}) and~(\ref{s&1/2_psi}), we have
\begin{eqnarray}
|\psi\rangle_{1,\frac{1}{2}}&=&\left(C_0^{(\frac{3}{2})}\left|\frac{3}{2},-\frac{3}{2}\right\rangle+C_1^{(\frac{3}{2})}\left|\frac{3}{2},-\frac{1}{2}\right\rangle+C_2^{(\frac{3}{2})}\left|\frac{3}{2},\frac{1}{2}\right\rangle\right.\nonumber\\
  &&\left.+C_3^{(\frac{3}{2})}\left|\frac{3}{2},\frac{3}{2}\right\rangle\right)+\left(C_0^{(\frac{1}{2})}\left|\frac{1}{2},-\frac{1}{2}\right\rangle+C_1^{(\frac{1}{2})}\left|\frac{1}{2},\frac{1}{2}\right\rangle\right),\label{1&1/2_total_state}
\end{eqnarray}
where $C_{n}^{(\frac{3}{2})}$ and $C_{n}^{(\frac{1}{2})}$ denote the coefficients of the $j=3/2$ and $j=1/2$ case, respectively.
%where $C_{n}^{(3/2)}$ and $C_{n}^{(1/2)}$ denote the coefficients of the $j=3/2$ and $j=1/2$ case, respectively.

From Eq.~(\ref{C_n&z}), we obtain
\begin{eqnarray} 0&=&C_0^{(\frac{3}{2})}z^0_{\frac{3}{2}}-\sqrt{3}C_1^{(\frac{3}{2})}z^1_{\frac{3}{2}}+\sqrt{3}C_2^{(\frac{3}{2})}z^2_{\frac{3}{2}}-C_3^{(\frac{3}{2})}z^3_{\frac{3}{2}},\label{s-1/2_mutipelt}\\
0&=&C_0^{(\frac{1}{2})}z^0_{\frac{1}{2}}-C_1^{(\frac{1}{2})}z^1_{\frac{1}{2}},\label{s-1/2_singlet}\\
z_{1,\frac{1}{2}}&=&\frac{C_{1-\frac{1}{2}}}{C_{1+\frac{1}{2}}}=\frac{\sqrt{|C^{(\frac{1}{2})}_{0}|^2+|C^{(\frac{1}{2})}_{1}|^2}}{\sqrt{|C^{(\frac{3}{2})}_{0}|^2+|C^{(\frac{3}{2})}_{1}|^2+|C^{(\frac{3}{2})}_{2}|^2+|C^{(\frac{3}{2})}_{3}|^2}},
\end{eqnarray}
where $z_{j}$, $z_{1,1/2}$ denote the characteristic variables of the spin-$j$ case and the pseudo spin case, respectively, and we also have the normalization condition $|C_{1+1/2}|^2+|C_{1-1/2}|^2=1$.

Solving Eqs.~(\ref{s-1/2_mutipelt}) and~(\ref{s-1/2_singlet}), we can easily gain the whole MSR of the spin-$s$ and spin-$1/2$ particles system with Eq.~(\ref{z_tan&e}). To illustrate the idea above, we show the MSR of the spin-$s$ and spin-$1/2$ particles system with time evolution, the real phase parameter $\varphi$ and $\delta=0$.

\begin{figure}[htbp]
\centering
\begin{minipage}{1\linewidth}
\centering
\begin{overpic}[width=0.667\linewidth,height=0.667\linewidth]{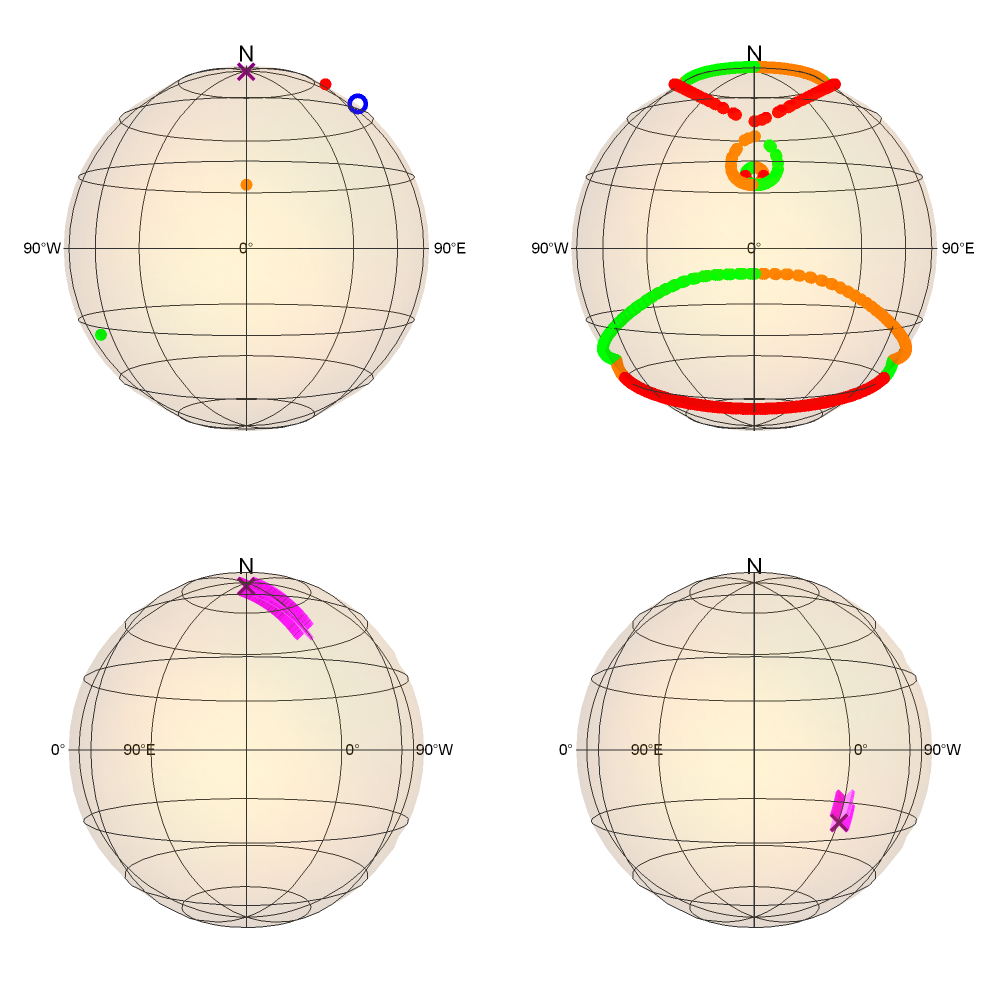}%\label{fig3-a}
\put(0,90){(a)}
\put(50,90){(b)}
\put(0,45){(c)}
\put(50,45){(d)}
\end{overpic}
\end{minipage}
\protect\caption{\label{fig2}(color online) Bloch representation of the spin-$1$ and spin-$1/2$ particles system with time evolution $t\in[0,4\pi]$, fixed real phase parameter $\varphi$ and $\delta=0$. a) All stars with $\varphi=\pi/6$ when time fixed at $t=\pi/4$.
b) The analogous mutiplet stars with the phase parameter fixed at $\varphi=\pi/6$. c) and d) The stars of the pseudo spin at $\varphi=\pi/4$ and $\varphi=5\pi/4$. The red dots, green dots and orange dots represent the stars of the analogous multiplet, blue circles represent the stars of the analogous singlet, and purple crosses represent the stars of the pseudo spin.
%Bloch representation of the spin-$s$ and spin-$1/2$ particles system with time evolution $t\in[0,4\pi]$, fixed real phase parameter $\varphi$ and $\delta=0$. \subref{fig2-c} and \subref{fig2-d} All stars with $\varphi=\pi/6$ and $\varphi=2\pi/3$. \subref{fig2-c} and \subref{fig2-d} The right views of the bloch spheres that contain the analogous mutiplet stars and analogous singlet stars, meanwhile, the phase parameter fixed at $\varphi=\pi/6$, respectively. \subref{fig2-e} and \subref{fig2-f} The stars of the pseudo spin at $\varphi=\pi/4$ and $\varphi=5\pi/4$.
}
\end{figure}

Figure \ref{fig2} shows the stars of the spin-$s$ and spin-$1/2$ particles system with time evolution, fixed real phase parameter $\varphi$ and $\delta=0$.
As shown in Fig.~\ref{fig2}(a), we can use three (one and one) sets of stars to represent the analogous mutiplet (the analogous singlet and the pseudo spin, respectively).
Moreover, focusing on one period of time, all stars of the mutiplet will jointly form a pattern that is plane symmetry around the prime meridian plane (Fig.~\ref{fig2}(b)).
However, unlike other sets of stars, the stars of the pseudo spin are mapped to the prime meridian and have the northernmost (near the north pole) and southernmost position at $\varphi=\pi/4+2k\pi$ with $t=n\sqrt{2}\pi$ and $\varphi=5\pi/4+2k\pi$ with $t=(2n+1)\pi/\sqrt{2}$ ($k,n$ are integers), respectively (Figs.~\ref{fig2}(c) and~\ref{fig2}(d)). %As shown in \cref{fig3-e,fig3-f}, the mutiplet stars can form legible and beautiful symmetry patterns.
%have the northernmost (the north pole) and southernmost position at $\varphi=\pi/4+2k\pi$ with $t=n\sqrt{2}\pi$ and $\varphi=5\pi/4+2k\pi$ with $t=(2n+1)\pi/\sqrt{2}$ ($k,n$ are integers), respectively [\cref{fig2-e,fig2-f}]. %As shown in \cref{fig3-e,fig3-f}, the mutiplet stars can form legible and beautiful symmetry patterns.
It means that the proportion of $|\psi\rangle_{s-1/2}$ has a lower bound and can asymptotically approach $100\%$ at specific time and phase parameters.

\begin{figure*}[htbp]
\centering
\begin{minipage}{\linewidth}
\centering
\begin{overpic}[width=1\linewidth,height=0.667\linewidth]{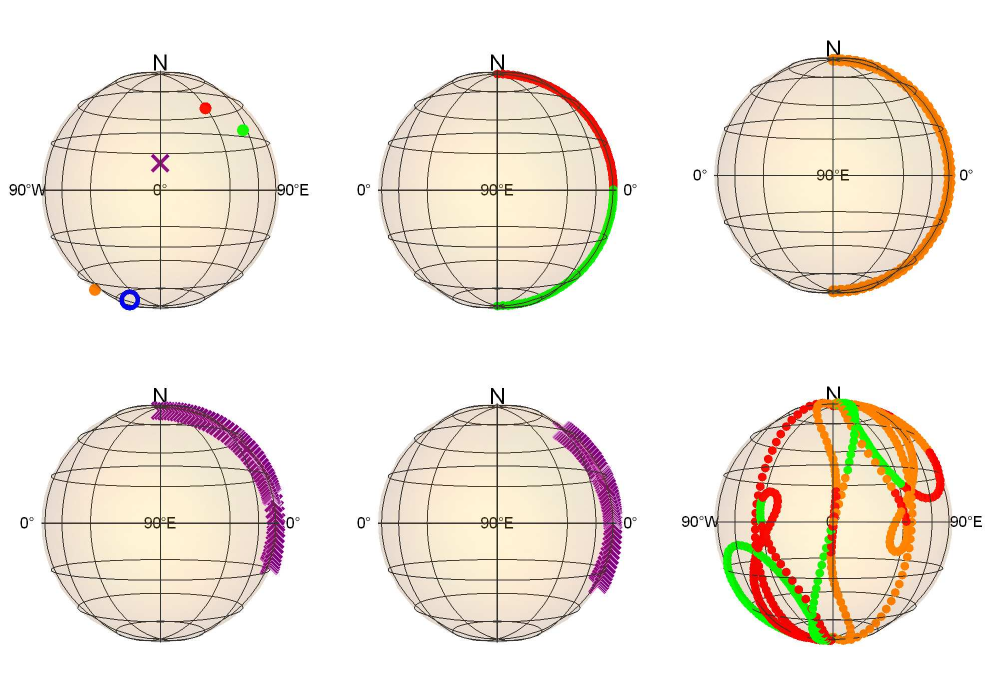}%\label{fig3-a}
\put(0,60){(a)}
\put(35,60){(b)}
\put(70,60){(c)}
\put(0,30){(d)}
\put(35,30){(e)}
\put(70,30){(f)}
\end{overpic}
\end{minipage}
\protect\caption{\label{fig3}
(color online) Bloch representation of the spin-$1$ and spin-$1/2$ particles system with the real phase parameter $\varphi$ variety, fixed time and $\delta=0$. a) All stars with $\varphi=3\pi/4$ and $t=\pi/2$. b-d) Two sets of the analogous mutiplet stars, another set of the analogous  mutiplet stars and the stars of the pseudo spin are mapped to the Bloch sphere with the variety of $\varphi$ and time fixed at $t=0$, respectively. e) The stars of the pseudo spin with $t=\pi/\sqrt{2}$. f) The analogous mutiplet stars with $t=\pi/3$.
The red dots, green dots and orange dots represent the stars of the analogous multiplet, blue circles represent the stars of the analogous singlet, and purple crosses represent the stars of the pseudo spin.
%Bloch representation of the spin-$s$ and spin-$1/2$ particles system with the real phase parameter $\varphi$ variety, fixed time and $\delta=0$. ($a$) All stars with $\varphi=3\pi/4$ and $t=\pi/2$. ($b$), ($c$), ($d$), ($e$) Two sets of the analogous mutiplet stars, another set of the analogous  mutiplet stars, the analogous singlet stars and the stars of the pseudo spin are mapped to the Bloch sphere with the variety of $\varphi$ and time fixed at $t=0$, respectively. ($f$) The stars of the pseudo spin with $t=\pi/\sqrt{2}$. ($g$), ($h$) The right views of the bloch spheres that only contain the analogous mutiplet stars and the analogous singlet stars with $t=\pi/3$, respectively. The red dots, green dots and orange dots represent the stars of the analogous multiplet, blue circles represent the stars of the analogous singlet, and purple crosses represent the stars of the pseudo spin.
}
\end{figure*}

Figure~\ref{fig3} shows the stars of the spin-$s$ and spin-$1/2$ particles system with the real phase parameter $\varphi$, fixed time and $\delta=0$.
As shown in Fig.~\ref{fig3}(a), we can use three (one and one) sets of stars to represent the analogous mutiplet (the analogous singlet and the pseudo spin, respectively).
Without time evolution, two sets of the analogous mutiplet stars are mapped to the quarter of the prime meridian (Fig.~\ref{fig3}(b)).
Meanwhile, the other set of the analogous mutiplet stars is mapped to the half prime meridian (Fig.~\ref{fig3}(c)).
However, the stars of the pseudo spin can not form the half prime meridian, completely (Fig.~\ref{fig3}(d)).
Besides, the result is exactly the same with the result with $\delta=1$ and $t=0$ case, in other words, it is independent of time evolution when $\delta=1$ because of the symmetry of the three spin directions.
As shown in Figs.~\ref{fig3}(d) and~\ref{fig3}(e), the stars of the pseudo spin have the northernmost (the north pole) and southernmost position at $\varphi=\pi/4+2k\pi$ with $t=n\sqrt{2}\pi$ and $\varphi=5\pi/4+2k\pi$ with $t=(2n+1)\pi/\sqrt{2}$ ($k,n$ are integers), respectively.
It means that the proportion of $|\psi\rangle_{s-1/2}$ has a lower bound and can asymptotically approach $100\%$ at specific time and phase parameters.
%$\varphi=\pi/4+2k\pi$ with $t=n\sqrt{2}\pi$ and $\varphi=5\pi/4+2k\pi$ with $t=(2n+1)\pi/\sqrt{2}$ ($k,n$ are integers), respectively.
Furthermore, we find that all stars of the analogous mutiplet jointly form closed curves that are $180^{\circ}$ rotational symmetric around the $x$-axis (the intersecting line of the prime meridian plane and the equatorial plane) (Fig.~\ref{fig3}(f)).

The method we proposed can visualize the mixed-spin ($s$, $1/2$) system and describe the entanglement of the system.
For the mixed-spin ($1$, $1/2$) system, $\left|2\right\rangle_1\left|1\right\rangle_2$ ($\left|0\right\rangle_1\left|0\right\rangle_2$) only belongs to the spin-($s+1/2$) sector.
Although $\left|1\right\rangle_1\left|0\right\rangle_2$ ($\left|0\right\rangle_1\left|1\right\rangle_2$, $\left|1\right\rangle_1\left|1\right\rangle_2$, $\left|2\right\rangle_1\left|0\right\rangle_2$) exists on both sectors and it cannot be separated into one sector, its stars are respective on the south pole and the equator in MSR.
The special entangled states
%$\left|\Psi^\pm\right\rangle=\frac{1}{\sqrt{2}}\left(\left|01\right\rangle\pm\left|01\right\rangle\right)$ and $\left|\Phi^\pm\right\rangle=\frac{1}{\sqrt{2}}\left(\left|00\right\rangle\pm\left|11\right\rangle\right)$
$\left(\left|1\right\rangle_1\left|0\right\rangle_2\pm\left|0\right\rangle_1\left|1\right\rangle_2\right)/\sqrt{2}$ and
$\left(\left|2\right\rangle_1\left|0\right\rangle_2\pm\left|1\right\rangle_1\left|1\right\rangle_2\right)/\sqrt{2}$
only belong to one sector and their stars on the south pole or the equator in MSR.
However, the stars' distributions of other entangled states are very complex.
Therefore, it is easy to find that all the non-entangled states' stars are on the south pole and the equator in MSR, so are special entangled states. But the distributions of general entangled states' are complex.

\section{Conclusion}\label{sec5}
Recently, the Majorana's stellar representation and relevant applications have demonstrated that the distributions and motions of the Majorana stars on the Bloch sphere have become a new and effective tool to study the symmetry-related questions in the high-dimensional or many-body system. Our study here shows that, utilizing the MSR, an arbitrary pure state can be always represented on a Bloch sphere with $4s+1$ stars in a two-spin ($s$ and $1/2$) system. We take the system described by coupling bases as a state of a pseudo spin-$1/2$. Furthermore, we propose a practical method to decompose the arbitrary pure state that can be regarded as a state of a pseudo spin-$1/2$.

As we know, a star on a Bloch sphere can represent a pure state, and a set of stars on a Bloch sphere can represent a high-dimensional pure states. However, this method can not be applied in arbitrary high-dimensional mixed states. Our result is not easy to be generalized to arbitrary high-dimensional mixed-spin systems, such as a mixed-spin $(s, 1)$ system. Because we can not directly use an effective pseudo spin $1$ to describe it.

Taking advantage of the MSR, which provides an intuitive way to study high spin system from geometrical perspective, we can have a novel and holonomic view to investigate intricate system.
Considering spin-$(1/2, 1/2)$ and spin-$(1, 1/2)$ systems, we find that one can easily distinguish between non-entangled states and entangled states by the distribution of stars in MSR.
Since we have presented a general method to visualize the mixed-spin $(s, 1/2)$ systems, one can further apply the method in many situations, such as quantum transitions and quantum tunneling in quantum spin baths and quantum dots.

\addcontentsline{toc}{chapter}{Acknowledgment}
\section*{Acknowledgment}
We thank Jing Cheng for useful discussions.

\addcontentsline{toc}{chapter}{References}
%\bibliographystyle{alpha}
%\bibliography{CPBRef}

%\end{CJK*}  %% end the Chinese environment
%\end{document}  %%% end document
\newpage

%\end{CJK*}  %% end the Chinese environment
\end{document}